# The Classification of M 1-78

G. T. Gussie

Department of Physics

University of Tasmania

GPO Box 252C

Hobart, Tasmania

Australia,

7001

internet: Grant.Gussie@phys.utas.edu.au

Abstract

The published properties of M1-78 are discussed with the purpose to resolve the object's classification as either a planetary nebula or an ultracompact H$_{II}$ region. A classification as a planetary nebula is rejected primarily because of the high luminosity of the object, but because of the chemical composition and expansion velocity of the nebula, a novel classification is proposed instead: that of an ultracompact H$_{II}$ region with a post-main sequence central star (possibly a WN star). It must therefore follow that observable ultracompact HII regions persist beyond the main sequence lifetimes of at least some massive stars, and so cannot be transient phenomena that are seen only during pre-main sequence or early main sequence evolution.

1) Introduction

The nebula M1-78 is an interesting and enigmatic object which has been the subject of optical, radio, and submillimetre observations for several decades. However its classification still remains undetermined. It has been classified as both an ultracompact (hereafter UC) HII region and a planetary nebula (hereafter PN). The UC HII regions are believed to be the products of massive star formation, and result when hot young stars photo-ionise their natal molecular gas clouds. A PN is thought to result when a more moderately massed star reaches the end of its main-sequence and red giant life, ascends the asymptotic giant branch (hereafter AGB), and then sheds its atmosphere to surrounding space to reveal its hot core. Ultracompact HII regions and PNe are therefore very different in their respective origins, but perhaps surprisingly they are not always easily distinguishable observationally. This paper discusses their distinguishing characteristics and the evidence for the classification of the object M1-78.

2) Evidence for the Classification of M1-78.

One distinguishing characteristic between a planetary nebula and a compact HII region is that the mean dust temperature of PNe is considerably higher than the mean dust temperature near UC HII regions. As a result of this, PNe and UC HII regions occupy different regions of a far-infrared colour-colour diagram. M1-78's far infrared colour indexes are

$$\frac{F_\nu(12\mu m)}{F_\nu(25\mu m)} = 0.22$$

$$\frac{F_\nu(25\mu m)}{F_\nu(60\mu m)} = 0.30$$

and therefore the nebula occupies a position on the infrared colour-colour plot in the transition region between planetary nebulae and compact HII regions (figure 1). Thus neither classification is clearly indicated, although possibly the planetary nebula classification is weakly favoured.

Another distinguishing characteristic is that (as a whole) compact HII regions tend to be more closely confined to the Galactic plane. Almost all UC HII regions in our Galaxy lie within 2° of the Galactic plane, and the overwhelming majority lie within 1° of the Galactic plane (Churchwell 1990). Planetary nebulae have a much wider distribution in Galactic latitude, reflecting their mixed Population I and Population II origins, but over 15% of catalogued planetary nebulae are within 2° of the Galactic plane (from Perek and Kohoutek 1965). M1-78 is a Galactic plane object, having a latitude of $b^{II} = 1° 28' 29"$. This would argue weakly that it is more likely to be an UC HII region than a PN, but it is obviously inconclusive.

Ultracompact HII regions are often less regular in appearance than planetary nebulae, which are usually either spheroidal shells or bilobate. In contrast UC HII regions show a variety of morphologies, including a significant percentage of irregular and "cometary" shapes. Only 4% of compact HII regions have "shell-type" morphologies — the classical PNe morphology — consisting of a dense outer shell surrounding a central hollow (Wood and Churchwell 1989). There are however many UC HII regions that are either spheroidal or unresolved ($\approx$ 60% of examples, Wood and Churchwell 1989), and there are also a few distinctly irregular planetary nebulae (*e.g.* NGC 7027). The morphology of M 1-78 has been determined by Scott and Harris (1978) who find that the nebula has bipolar bright region with a size of 4.0" × 1.8" aligned along a north-west to south-east line, as would be fairly typical of a small planetary nebula. However, they also find an irregular fainter region extending some 4" to the north-east, which would seem more typical of an UC HII region. Scott and Harris find that neither classification can be ruled out based on the observed morphology.

Ultracompact HII regions generally occur in or near active star forming regions, and consequently are often seen in groups. The stellar progenitors of PNe are aged stars that have wandered far from their natal clouds, and therefore are usually isolated or lie within aged clusters. A radio survey of the region surrounding M 1-78 by Scott and Harris (1978) does not show any nearby HII regions that can clearly be associated with M 1-78. This would argue weakly that the nebula has probably left its formation region and is therefore likely to be a PN. But alternatively, M 1-78 may simply be the only massive star to form in that particular cloud.

Ultracompact HII regions have fairly uniform chemical compositions reflecting the chemical composition of the general interstellar medium (hereafter ISM). Planetary nebulae often have significantly modified chemical compositions as a result of the convective mixing of nuclear-processed material into the outer layers of their precursor stars. Zijlstra *et al.* (1990) note that the chemical composition of M 1-78 is enriched in nitrogen with respect to oxygen when compared to typical HII regions, although no firm numbers are given. Perinotto (1991) gives the nitrogen to oxygen abundance ratio of 0.47 for M 1-78 — near the average value for his PNe but over twice the canonical N:O abundance ratio of 0.22 for the ISM (Mezger 1972). This significant deviation from the chemical composition of the ISM would favour identification of M 1-78 as an evolved object such as a PN.

Ultracompact HII regions often have lower expansion velocities than planetary nebulae. Typically, compact HII regions have ionized-gas expansion velocity less than or equal to the sound speed in their gas, that is $v_{exp} \lesssim 10$ km s$^{-1}$ Exceptions to this occur when the ionizing star itself has a supersonic velocity of >10 km s$^{-1}$ with respect to its natal nebula, as occurs in multiple-star OB associations. In such cases, bow shocks develop that distort the HII region into a cometary structure (Mac Low *et al.* 1991). Planetary nebulae as a group have somewhat higher measured ionized-gas expansion velocities, averaging $v_{exp} \approx 20$ km s$^{-1}$ (Weinberger 1989), and as such *typically* expand supersonically, even those with a regular shell-type morphologies. The measured expansion velocity of M1-78 is $v_{exp} = (25.3 \pm 2.0)$ km s$^{-1}$ (Gussie and Taylor 1989). This supersonic expansion velocity favours classification as a PN since the morphology of the nebula is not cometary, and displays no other manifestations of a bow shock created by a supersonic central star.

The active star forming regions that compact HII regions are generally associated with clouds of molecular gas which are observable as CO emission (*e.g.* Churchwell *et al.* 1992). In contrast, the majority of planetary nebulae show a surprising paucity of CO emission, although some are known to be relatively weak CO sources (*e.g.* Huggins and Healy 1989). M1-78 is a powerful source of CO emission, and the only known PN with a comparable intensity of CO emission is NGC 7027. There are however considerable differences between the CO clouds of NGC 7027 and M1-78. Zuckerman *et al.* (1977) report an extended distribution for the CO of M1-78, while the CO emission of NGC 7027 is confined to within one arc minute of the center of the ionized nebula (Phillips *et al.* 1991). Bujarrabal *et al.* (1988) report that the linear diameters of CO clouds of various young planetary nebulae and proto-planetary nebulae is 0.23 pc, to within a factor of 2. A linear size of 0.23 pc for the CO cloud of M1-78 would require the maximum distance to M1-78 be an improbably low 260 pc. This would imply that the CO in M1-78 is associated with an external molecular cloud, and so a classification of the nebulae as an UC HII region would be supported.

The CO emission lines M1-78 also have a very narrow bright component and a broader low-intensity component (figure 2), while the CO emission lines of planetary nebulae are typically parabolic, flat-topped, or double-peaked in shape (*e.g.* Huggins and Healy 1989). Furthermore, the systemic velocities of the CO lines of NGC 7027 closely match the velocity of its optical emission lines, while the velocities of the molecular and optical emission lines are displaced by $\approx 10$ km s$^{-1}$ in the case of M1-78. This would favour identification as an UC HII region because the molecular emission lines of UC HII regions are often shifted with respect to the optical emission lines by a similar velocity (Forster *et al.* 1990).

An alternative explanation for the strange CO emission properties of M1-78 is that it may be a planetary nebula that possesses a CO maser in its molecular envelope. Spherically-symmetric radiative transfer models of the expanding neutral envelopes of AGB stars by Morris (1980) have shown that under certain fairly narrowly defined conditions of mass-loss rate and expansion velocity, CO maser emission may occur. Morris identified the protoplanetary nebula CIT 6 as undergoing possible CO maser emission because it has displayed changes in the shape and intensity of its $J = 1 \rightarrow 0$ CO emission line over a period of about two years, which Morris attributes to the switching off of a weak CO maser. An obvious objection to the hypothesis that M 1-78 has a CO maser is that the CO maser emission in the Morris' models are double-peaked while M 1-78 displays only a single red shifted peak. However, the Morris' models are spherically symmetric and therefore have identical conditions in both the near (blue shifted) and far (red shifted) sides of their envelopes. It is plausible that the lop-sided CO emission observed in M 1-78 is the result of an asymmetric mass distribution. The apparent rapid shut off of the CO maser in CIT 6 illustrates how sensitive CO maser emission is to density and temperature changes, making it reasonable that CO maser emission could exist in only one side of a circumnebular envelope. This hypothesis could however not easily explain the velocity of the observed H I absorption line (Gussie and Taylor 1994), which is identical to that of CO emission. The presence of H I absorption would indicate that the bulk of the red shifted neutral gas is between us and the ionized nebula, and must therefore be either unrelated to the ionized nebula or undergoing infall rather than the outflow expected of a PN molecular shell. The CO emission of M 1-78 would therefore tend to favour a classification as an UC H II region.

The luminosity of the nebula would also tend to argue against its identification as a PN. Based on far-infrared data taken by the Infrared Astronomy Satellite, Puche *et al.* (1988) find a total far-infrared luminosity of $L_* = \frac{D}{6\,\text{kpc}} 10^5 L_\odot$, where $D$ is the nebula's distance in kiloparsecs. If the range in planetary nebula nucleus masses is taken to be 0.5 $M_\odot$ to 0.8 $M_\odot$ (Pauldrach *et al.* 1988), then the theoretical upper limit to the luminosity of PNe is $\approx 20000 L_\odot$ based on models of Paczynski (1970) and Kippenhahn (1981). An upper limit to the distance of M 1-78 would be $D = 1.2\,\text{kpc}$ if it were a PN. However, M 1-78 suffers a total extinction of $A_{H\beta} = 10.2$ magnitudes (Scott and Harris 1978), while Allen (1973) list the mean absorption of optical radiation in the Galactic plane as $A = 1.9$ magnitudes kpc$^{-1}$. M 1-78 and its surrounding environs must therefore provide an additional optical extinction of $A_{H\beta} = 8.1$ magnitudes if the nebula were a PN at 1.2 kpc distance. This is very high, but not entirely unprecedented. Woodward *et al.* (1992) have shown that the optical extinction of NGC 7027 (while very patchy) reaches $A \approx 8$ magnitudes at least at some points on the face of the nebula's ionized region. Various proto-planetary nebulae also show high optical extinctions (*e.g.* Soptka *et al.* 1985). If M 1-78 were a PN it would then seem to be a very young, massive, and dusty object, perhaps the most massive example known. Its properties would perhaps place it as an intermediate object in evolutionary terms between the massive PN NGC 7027 and the proto-planetary nebula CRL 618, which possesses a massive and dusty molecular envelope (Bunjarrabal *et al.* 1988) but not as yet as highly a developed ionised region as M 1-78. Alternatively, the nebula may be a more ordinary PN that happens to lie behind a dense molecular cloud; at least a few such juxtapositions must exist. However compact HII regions are *typically* located near (or within) cold molecular clouds and often suffer from tremendous optical extinction. In fact, most UC H II regions are completely obscured at optical wavelengths and are only seen as radio and infrared sources.

Ultracompact HII region regions are young objects, and as such they are assumed to have much smaller peculiar velocities than do PNe; that is, their motion is expected to follow the Galactic rotation curve more closely. Kinematic distances are therefore routinely used to estimate the distances to UC HII regions, but they have long since been proven to be unreliable in determining distances to PNe. Therefore, if M 1-78 were shown to have a distance significantly different from its kinematic value, identification as a PN would be indicated. The systemic velocity of the ionized nebula of M 1-78 (Gussie and Taylor 1989) is found to be $v_{LSR} = (-76 \pm 2)\,\mathrm{km\,s^{-1}}$ as measured by optical spectroscopic observations of the [OIII] lines. The optical velocity is supported by measurements of hydrogen radio recombination lines by Terzian *et al.* (1984) and Churchwell, Terzian, and Walmsley (1976). This would imply a kinematic distance to the nebula of $D \approx 8\,\mathrm{kpc}$.

Comparison of the kinematic distance with the actual distance is however difficult, since the distance to M 1-78 is poorly known. Various statistical PN distance determination methods have given distances from 3 kpc to 5 kpc (Cahn and Kaler 1971, Acker 1978, Maciel 1984), but the reliability of these methods has not been shown and is in considerable doubt; and they would of course be inapplicable anyway if M 1-78 were not a PN.

More reliable attempts to determine the distance to M 1-78 include that by Puche *et al.* (1988), who present $\lambda = 21\,\mathrm{cm}$ spectral line observations which show HI absorption features at $v_{LSR} = -10\,\mathrm{km\,s^{-1}}$, $v_{LSR} = -45\,\mathrm{km\,s^{-1}}$, and $v_{LSR} = -65\,\mathrm{km\,s^{-1}}$. Puche *et al.* identify the $v_{LSR} = -10\,\mathrm{km\,s^{-1}}$ absorption feature as interstellar absorption caused by HI the Local Spiral Arm and the $v_{LSR} = -45\,\mathrm{km\,s^{-1}}$ absorption feature as interstellar absorption caused by HI in the Perseus Arm. The $v_{LSR} = -65\,\mathrm{km\,s^{-1}}$ absorption feature is tentatively identified as being due to a third spiral arm beyond the Perseus Arm. The presence of the HI absorption would require the nebula to be between 6 kpc and 8 kpc, in accord with the kinematic distance. This distance would then preclude the possibility that M 1-78 is a planetary nebula on the basis of its impossibly high required luminosity. This distance would also place the nebula $\approx 200$ pc above the Galactic plane, which is an unusually large, but not an impossibly large, height for an UC HII region.

The distance might be determined by further high-resolution radio continuum observations. If the nebula were a PN (and therefore at a distance of $D = 1.2\,\mathrm{kpc}$) then the rate of angular growth of the ionized region would be $\frac{\Delta \theta}{\Delta t} = 0.0045''\,\mathrm{yr^{-1}}$. A high-resolution $\lambda = 2\,\mathrm{cm}$ continuum image of the nebula taken at the present epoch could therefore be used to measure the apparent expansion by comparison with the image taken by Scott and Harris in 1978.

The uncertainty regarding the nebula's luminosity could be answered if its central star (or stars) was observed. However, the author knows of no observations or detections of the nebula's central star, and the large external and internal dust extinction makes it unlikely that the star's optical emission would be detectable.

4) Conclusions

It is clear that the various distinguishing characteristics of planetary nebulae and ultracompact HII regions do not give an unambiguous classification for M 1-78. On the whole, a classification as an UC HII region would seem to be favoured but the unusual chemical composition of the nebula and its high expansion velocity would present serious problems for such a classification. I consequently favour a rather unique classification for this object: an UC HII region with a post-main sequence central star.

Wood and Churchwell (1990) have shown that UC HII regions can not be transitory phenomena that are created at the start of main sequence evolution as was previously assumed, but must instead persist for much of a massive star's main sequence life time. They propose and examine many possible mechanisms for the preservation of an identifiable UC HII region long past the time that a freely expanding UC HII region would require to disperse into the ISM. If an UC HII region persisted for even longer time scales than those suggested by Wood and Churchwell, then a very aged UC HII region would receive chemical enrichment from a post-main sequence stellar wind. Such a wind would release nuclear-processed stellar material brought to the photosphere by convection during one or more "dredge-up" phases of stellar evolution. Such deep convective mixing is not expected until after main sequence evolution (Greggio 1983). The observed over-abundance of nitrogen in M 1-78 would suggest that the central star is (or was) a nitrogen-rich Wolf-Rayet star (spectral class WN). The evolutionary status of WN stars has been a matter of some debate (*e.g.* Underhill 1983), but they are now generally regarded as post-main sequence objects with a high original mass (*e.g.* Abbott and Conti 1987).

As a point of clarification, the central stars of several PNe are known to have Wolf-Rayet type stellar spectra (*e.g.* NGC 6369). It is however believed because of the differences in the distributions of PNe and OB associations that these "Wolf-Rayet" stars have quite different origins than those associated with young O stars, and are instead post AGB objects of much more moderate original mass. These PN Wolf-Rayet stars are not considered here. Instead, it is suggested that the power source of M 1-78 is a "classical" Wolf-Rayet star, and as such is an extreme Population I object.

The high luminosity of the nebula is not a difficulty if a WN star were its power source, as the absolute magnitudes of such objects are as high as $M_v = -8$ (Abbott and Conti 1987). WN stars are also often seen to have close O-type binary companions (Roberts 1962), increasing the available luminosity even further.

It is known that WN stars are usually significantly less massive than their O star companions (Massey 1981), despite being the (apparently) more evolved of the pair. This is easily explained if much of the stellar mass is lost to the Wolf-Rayet stellar wind. Mass loss in Wolf-Rayet stars is observed at the $\approx 5\times 10^{-5}$ $M_\odot$ year$^{-1}$ rate (Abbott *et al.* 1986), which over the expected Wolf-Rayet lifetime of $\approx 4\times 10^5$ yr (Chiosi *et al.* 1978) would strip $\gtrsim 10$ $M_\odot$ from the star. Wolf-Rayet stars are therefore regarded as the bare cores left by the post-main sequence loss of the outer envelope of an O star (Maeder 1983). So-called "Ring Nebulae" have been observed around many Wolf-Rayet stars which are believed to consist of the remnant wind material augmented by swept-up interstellar matter. Chemical compositions of the nebulae around WN stars have shown nitrogen to helium abundance enhancements of 3 to 10 times over that of the general ISM (Kwitter 1981, Kwitter 1984, Esteban *et al.* 1990). The observed by nitrogen abundance anomaly in M 1-78 is therefore easily accounted for by the addition of a nuclear-processed stellar wind material to a pre-existing UC HII region.

WN star mass loss could also explain the supersonic expansion observed in M 1-78. Terminal wind velocities in Wolf-Rayet stars are observed to be as high as 3700 km s$^{-1}$ (Willis 1982), making the kinetic energy of Wolf-Rayet stellar winds a significant proportion of the stellar energy budget; perhaps as high as 10%. Such a wind would %%impact significantly on the kinematics of the surrounding gas. Kwok *et al.* (1978) have proposed an "interacting stellar winds" model for planetary nebula evolution, where a fast ($\gtrsim 2000$ km s$^{-1}$) wind from the PN central star overtakes the slower, relatively dense wind of AGB mass loss. This accelerates the ionised shell. Applying this same idea to an UC HII region with a WN central star, the star's fast wind would accelerate the gas of the natal molecular cloud, rather than material from an earlier mass-loss episode as in the case in PN evolution.

5) Summary

M 1-78's infrared colours, its isolation, its chemical composition, and its expansion velocity would argue that the nebula is a young but massive planetary nebula. However the molecular emission would favour a classification as an ultracompact HII region, and the nebula's luminosity makes M 1-78's classification as a planetary nebula untenable if the distance estimate of Puche *et al.* (1988) is accepted. The contradictory nature of the evidence concerning the classification of M 1-78 may be resolved if the nebula were an ultracompact HII region that has undergone significant chemical alteration by the stellar wind of a post-main sequence central star. The nebula's luminosity and enriched nitrogen abundance would argue that the central star is (or was) of spectral class WN. Observable ultracompact HII regions would then necessarily persist beyond the entire main sequence lifetime of a massive star.

6) Acknowledgments

This research was supported by a grant from the Australian Research Council. This research has made use of the SIMBAD database, operated by CDS, Strasbourg, France.

Figure Captions

1) IRAS Colour-Colour Plot of Planetary Nebulae and Compact HII Regions

IRAS colour-colour plot of a sample of planetary nebulae (circles) and compact HII regions (squares) as well as M 1-78 (cross) showing flux at $\lambda = 12\,\mu m$ over flux at $\lambda = 25\,\mu m$ in Jansky as a function of flux at $\lambda = 25\,\mu m$ over flux at $\lambda = 60\,\mu m$, also in Jansky. Planetary nebulae and compact HII regions occupy distinct regions on this plot, although some overlap occurs. M 1-78 has a position intermediate between the two classes of objects, but is perhaps more similar to the planetary nebulae. The planetary nebular data are from Pottasch *et al.* (1984) and consist of 41 bright planetary nebulae. The compact HII region data are from Antonopolou and Pottasch (1987) and consist of 64 nebulae.

2) $J = 3 \rightarrow 2$ CO Spectrum of M 1-78

The $J = 3 \rightarrow 2$ CO spectrum of M 1-78 from Gussie and Taylor (1994). The local standard of rest velocity of the ionized nebula is shown by the arrow. The molecular gas is red-shifted with respect to the ionized gas, and displays a highly-peaked line shape that is atypical of planetary nebulae. Note the highly red-shifted component near -7 $km\,s^{-1}$.

Figure 1

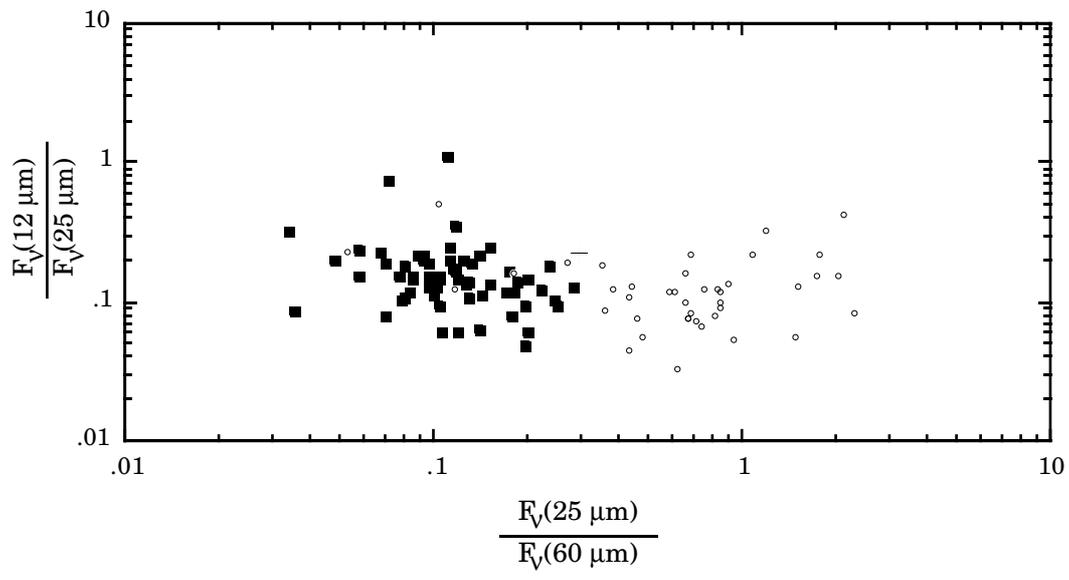

Figure 2

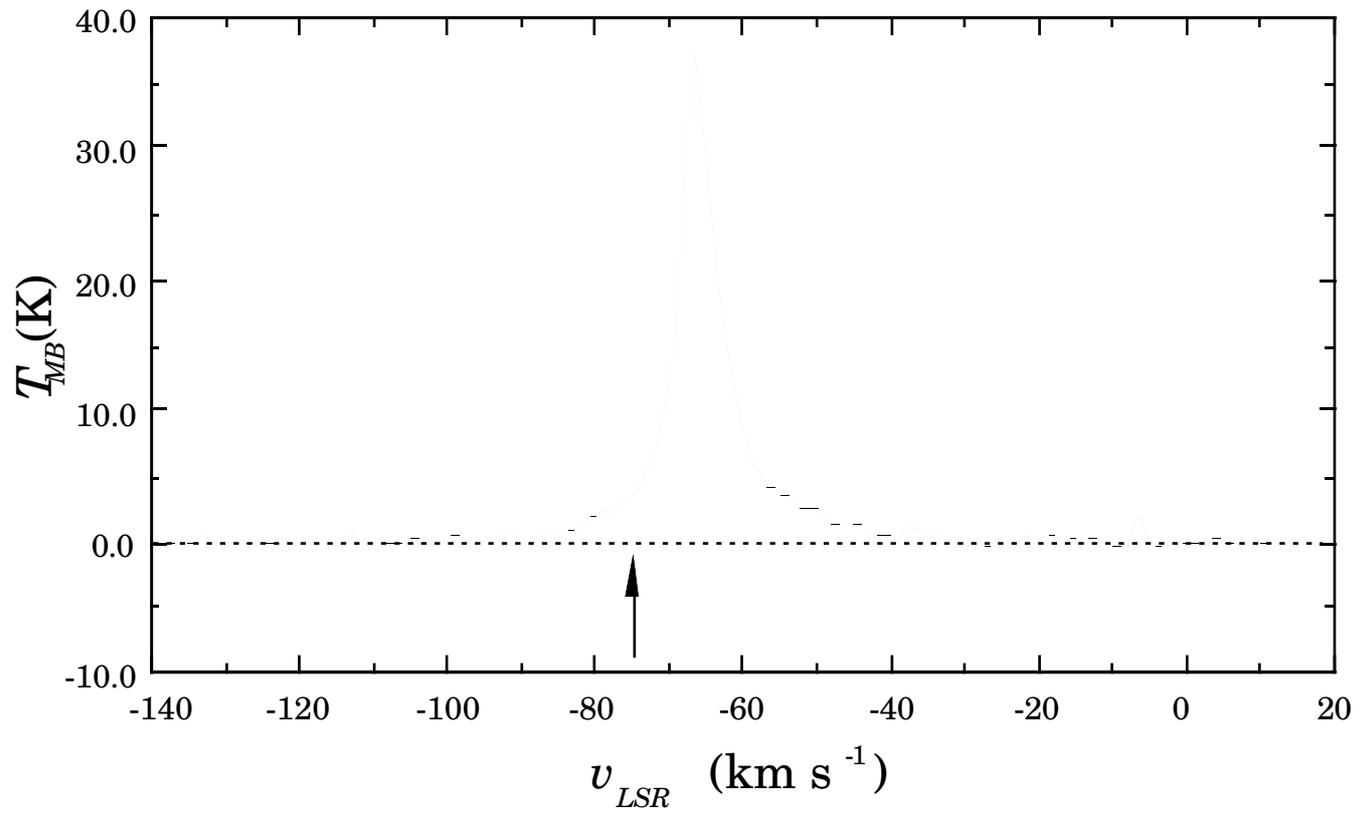